# Iceberg stability during towing in a wave field

Trygve K. Løken[1], Aleksey Marchenko[2], Jean Rabault[3,1], Olav Gundersen[1], Atle Jensen[1]
[1] University of Oslo, Oslo, Norway
[2] The University Centre in Svalbard, Longyearbyen, Norway
[3] Norwegian Meteorological Institute, Oslo, Norway

## ABSTRACT

Due to their large mass and small aspect ratio, icebergs pose a threat to boats and offshore structures. Small icebergs and bergy bits can cause harm to platform hulls and are more difficult to discover remotely. As icebergs are dynamic mediums, the study of icebergs in relation to safe human operations requires the rigorous analysis of the ice-ocean interaction, in particular with waves and currents. In this paper, we present iceberg towing experiments and analyze iceberg stability from GPS tracks and inertial motion unit data. The towline tension as well as the boat motion relative to the iceberg was measured. Different scenarios were investigated by changing the towing strategy with regards to towing speed, direction (straight or curved trajectory) and acceleration. Large amplitude roll oscillations with period of approximately 30 s were observed immediately after the load dropped and the iceberg returned to a stable static position. In two of the cases, the iceberg flipped over partly or entirely after some towing time. From the load cell, we observed oscillations in the system with periods of approximately 6 s, which were attributed to the rope elastic properties and the iceberg response. The load oscillations increased when the towing direction was against the waves as opposed to perpendicular to the waves.

## INTRODUCTION

The decline in the Arctic ice cover that has been observed over the past decades has allowed for more human activities in the region, such as shipping and exploitation of natural resources (Smith and Stephenson, 2013; Feltham, 2015). Even though the ice cover is decreasing, icebergs are observed in almost all Arctic seas (Abramov and Tunik, 1996). Drifting icebergs poses a threat to floating or fixed structures and their existence influences the concept of offshore development. Due to their large mass, icebergs can apply considerable pressure stress on platform hulls and cause mechanical failure. The technical term for small icebergs like the ones investigated in this study is bergy bit (< 1000 tons), but they will be referred to as icebergs in the text. Small icebergs can still influence damage on constructions, and they are more difficult to observe remotely, as pointed out by Marchenko et al. (2020).

When there is a risk of collision between icebergs and platforms, it is necessary to deflect its drifting course to ensure safe human operations in polar offshore regions. Iceberg towing is a well known technique and many experimental tests and studies have been performed, also in the resent years, e.g. Kornishin et al. (2019) and Efimov et al. (2019). Marchenko and Gudoshnikov (2005) identified several requirements for the towing operation to be successful, e.g. that the ship should have sufficient power to significantly change the iceberg trajectory, and the towline needs to resist the water and wave induced drag forces on the iceberg. Due to their small aspect ratio, especially spherical icebergs are prone to capsizing when external forces are applied. Such an event can cause harm to people and equipment during the operation. Knowledge of iceberg stability during towing is therefore of importance to reduce the risk of accidents (Marchenko, 2006).

Our aim with this study is to investigate iceberg stability during towing to improve the safety of such an operation. Unsuccessful towing events are often caused by towline slippage or iceberg overturning (Crocker et al., 1998). Therefore, we have applied sinking line towing, where the towline is partly submerged around the iceberg with the depth regulated with added buoyancy. This method reduces the probability of towline slippage. In addition, the towline force is applied closer to the iceberg center of rotation, which reduces the overturning moment and the risk of iceberg rollover (Crocker et al., 1998). Up until now, most studies on iceberg drift and towing have installed only GPS trackers on the icebergs, which give information about drift, but not on the iceberg dynamics. In this study, towing experiments with small icebergs are presented. The novelty of the investigation is the installation of Inertial Motion Units (IMUs) on the icebergs, which have allowed for observations of three-axes acceleration and rotation. Detailed surveillance of such motion, in addition to knowledge about the iceberg sub-surface geometry, which was obtained with a Remotely Operated Vehicle (ROV), gives insight in the hydrodynamic stability of the ice structures. Met-ocean parameters and the applied towing force were measured. The massive instrumentation allowed for the investigation of the effect of incoming waves and ocean current.

**DATA AND METHODS**

Towing experiments were carried out in the Tempelfjord near Longyearbyen on Svalbard in September 2020. Tunabreen glacier extends out in the fjord and frequent calving events makes it an ideal site for working with small icebergs and bergy bits. Figure 1 shows the location and a Sentinel-2 satellite image of the fjord and the glacier from September 22, 12:27 UTC.

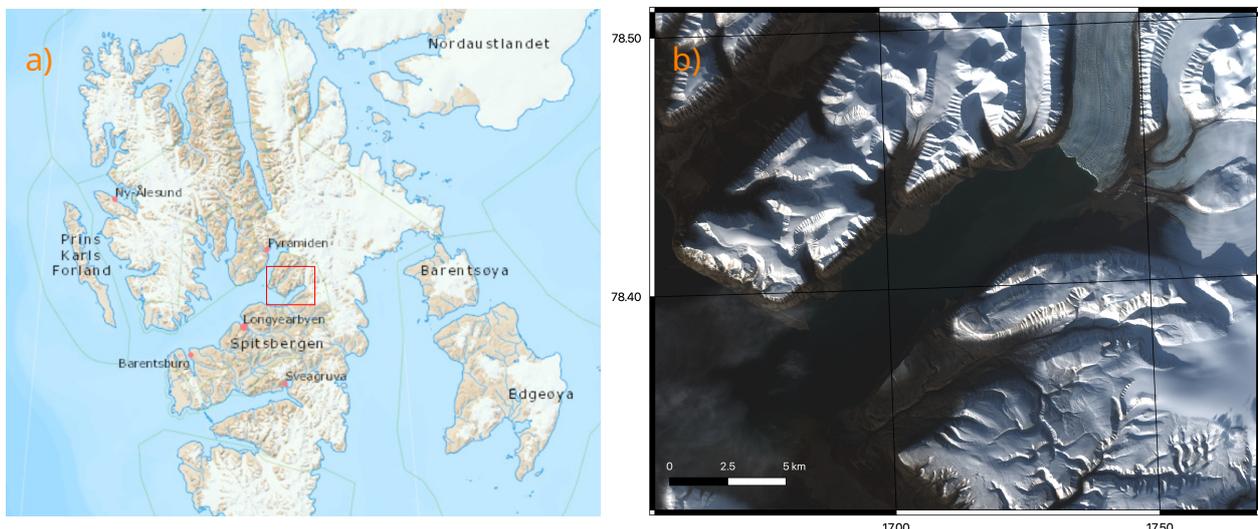

Figure 1. Location of the experiments. a) Map of the Svalbard Archipelago with the Tempelfjord indicated (TopoSvalbard, 2021). b) Satellite image of the Tempelfjord and the Tunabreen glacier.

**Experimental setup**

A 10.5 m long, 500 Hp Polarcircle 1050 boat with 7 tons total mass was used in the experiments. It was decided to use a fast boat in order to save time on the daily 50 km trip from Longyearbyen to Tempelfjord. Open source IMUs and GPS trackers (Rabault et al. 2017) were mounted on the boat roof and on the iceberg with ice screws to measure their absolute and relative motion. In addition, a pair of Garmin Astro GPS trackers were used as redundancy. The keel depth $h_k$ was determined and the sub-surface structure was investigated with an ROV. It was attempted to perform a video scan with the ROV to generate a 3D model of the sub-surface geometry, but the visibility was too poor due to the high concentration of glacier sediments. Wave motion was measured with an IMU and ocean current with an Acoustic Doppler Current Profiler (ADCP), which were mounted on buoys and moored to an anchor in the vicinity of the towing experiments.

After the installation of instruments on the iceberg and the buoys, the towing setup was arranged. A 12 mm thick polyester sinking rope was applied. The 92 m long rope was deployed around the iceberg from a small rubber boat while the large boat stayed at a fixed distance to the iceberg. Floaters on 1.5 m long straps were attached to the rope with 2-3 m spacing, in order to keep the towline submerged under the surface and prevent slippage during the operation. The rope ends were connected to a 5 tons load cell from Strainstall (type 12160-3). A 9 m long piece of double rope was attached to each side of the boat stern and connected to the load cell in the center. The joining point was kept floating with a buoy. The setup is illustrated in Fig. 2.

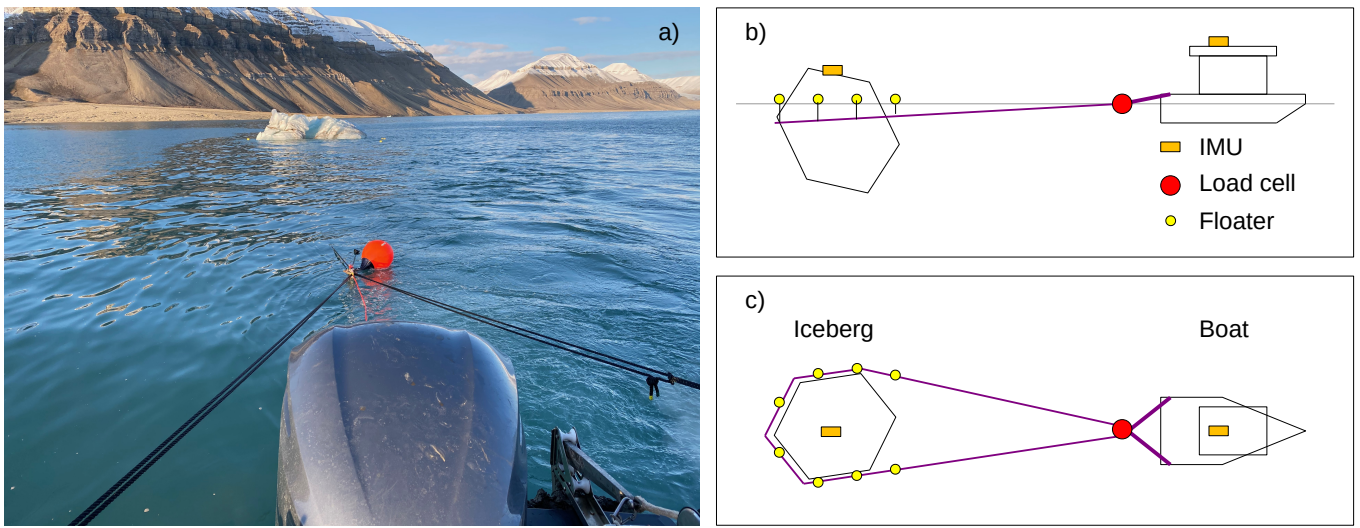

Figure 2. Experimental setup. a) Picture towards the iceberg from the boat stern. b) Towing sketch seen from the side. c) Towing sketch seen from above.

Three successful towing events were carried out and will be referred to as T1-T3. Experimental details are summarized in Table 1. At the end of T3, the iceberg rolled over and the IMU attached to it was lost, hence the missing parameters in Table 1. The iceberg area $S$ at the water line was estimated from a series of photos of the iceberg and an object of known dimensions from different angles. In T2, the photos were used to produce a 3D model of the iceberg above-surface geometry (not presented), in order to determine the area with better accuracy. The mean free board $h_{fb}$ was estimated from visual observations. The period of natural oscillation in heave $T_h$ was determined from the heave spectra from the IMU placed on the iceberg with the Welch method, described under "Data processing".

Table 1. Experimental details with emphasis on iceberg properties.

| Exp. | Date | Time (UTC) | $S$ [m²] | $h_k$ [m] | $h_{fb}$ [m] | $T_h$ [s] | $m_i$ [tons] |
|---|---|---|---|---|---|---|---|
| T1 | 22 | 14:20 ~ 15:30 | 31 | 8 | 1.6 | 7.32 | 424 |
| T2 | 24 | 11:40 ~ 12:20 | 28 | 5 | 0.8 | 4.69 | 158 |
| T3 | 25 | 11:40 ~ 12:00 | 29 | 4.5 | 1.0 |  | *145* |

Iceberg mass $m_I$ was estimated from $S$ and $T_h$ from the momentum equation in the vertical direction. Following Marchenko et al. (2020), inertial force is balanced by buoyancy and gravity forces

$$m_I \frac{d^2 z}{dt^2} = \rho_w g (V_W - Sz) - m_I g, \qquad (1)$$

where $z = z(t)$ is the vertical displacement of the iceberg (positive upwards) relative to the calm water surface at $z = 0$, $\rho_w$ is the water density, $V_W$ is the submerged iceberg volume in hydrostatic equilibrium and $g$ is the acceleration of gravity. In hydrostatic equilibrium, $\rho_w V_W = m_I$ and from the solution of Eq. 1, the iceberg mass is

$$m_I = \frac{\rho_w g S T_h^2}{4\pi^2}. \qquad (2)$$

In T3 when $T_h$ data were lost, $m_I$ was estimated from the approximated iceberg volume $S(h_{fb}+h_k)$ and a typical iceberg density of 910 kg/m$^3$ (Robe, 1980).

Different towing strategies were applied in the experiments, the boat heading was either straight or curved and the motor power was either constant or slowly increased. The different trajectories from the IMU GPS are presented in Fig. 3, where the boat position (blue) and the iceberg position during free drift before towing (gray) and towing (orange) are indicated. The starting time of the towing is indicated and 10 min intervals are marked with dots. Even though the instruments were lost when the iceberg tipped in T3, the Garmin Astro GPS tracker transferred coordinates via radio communication. The iceberg trajectory presented in Fig. 3c is therefore from the Garmin instrument. A comparison of the GPS tracks in T1-T2 (not presented) show good agreement between the two different instruments.

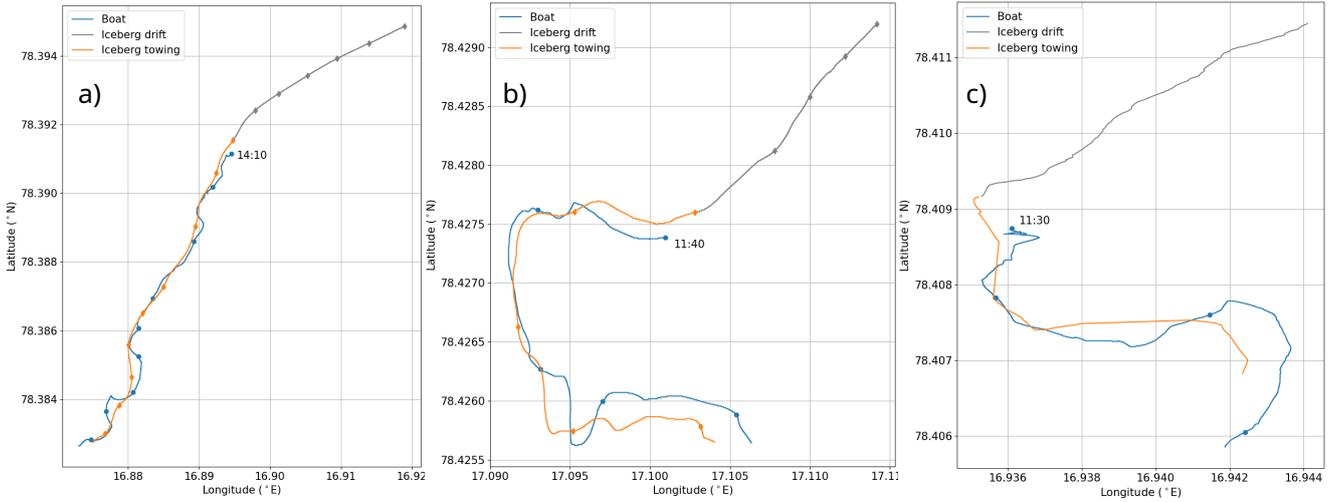

Figure 3. Boat and iceberg trajectory during towing and free drift. The starting time of the towing (in UTC) is indicated and 10 min intervals are marked with dots. a) T1. b) T2. c) T3.

Due to the elastic properties of the towline, oscillations can be expected in the iceberg-rope-boat system. Marchenko and Eik (2012) investigated the stability of steady towing, i.e. at constant towing speed, propulsion and water speed, by evaluating small fluctuations in the vicinity of the steady solution of the system's momentum equations in the axial direction. They estimated the period of the system oscillations $T_s$ and their Eq. (17) can be written as

$$T_s = 2\pi \sqrt{\frac{m_B}{dF_t/dX}}, \qquad (3)$$

where $m_B$ is the boat mass, $F_t$ is the rope tension, $X$ is the distance from the boat to the iceberg and the derivative is evaluated during steady towing. Equation 3 neglects the added mass of the boat

$m_{B,a}$ in the axial direction ( $m_{B,a}/m_B \ll 1$) and is valid when the boat mass is much smaller than the iceberg mass ($m_B/m_I \ll 1$), which was the case in all the present experiments.

**Data processing**

The sampling frequency was approximately 10 Hz for the IMUs and 1 Hz for the GPSs. Time series of speed, direction and distance between sensors were obtained from GPS positions. Signal smoothing of the GPS time series was performed with a third order polynomial Savitzky-Golay filter with a 41-point window size. Iceberg heave motion and ocean surface elevation were obtained from downward acceleration time series from the IMU placed on the iceberg and on the buoy, respectively. The time series were first re-sampled to obtain a constant sampling frequency of 10 Hz. Numerical integration of downward acceleration with respect to time was performed twice to obtain vertical displacement. After each integration step, a second order Butterworth bandpass filter with cutoff frequencies of 0.05 and 2 Hz was applied to remove any low frequency noise associated with the integration (Sutherland and Rabault, 2016). Displacement in the horizontal directions was obtained with the same approach.

Power Spectral Densities (PSDs) of displacement obtained from the double integrated accelerations, and of rotation obtained directly from the IMUs, were calculated from 30 min time series. The time series were subdivided into segments of 2048 data points with 50% overlap and the segments were Fourier transformed. The spectral estimates were ensemble averaged to decrease statistical uncertainties according to the Welch method, and a Hanning window was applied to each segment to reduce spectral leakage (Earle, 1996). The resulting PSD had approximately 25 degrees of freedom. Spectral 95% confidence intervals were estimated from the Chi-squared distribution (Earle, 1996).

The load cell output, which had a sampling frequency of approximately 10 Hz, was converted from voltage to kN from a calibration curve which was obtained by measuring the static load of three lifted objects with a known mass ranging 0.2-1.5 tons. Linear regression was applied to fit a straight line between the data points. A calibration was performed before and after the experiments, and the average slope was used.

**Oceanographic conditions**

Wave motion was measured in T2-3, but the IMU buoy was not deployed during T1. Significant wave height $H_S$ was found from $H_S = 4\sigma$, where $\sigma$ is the standard deviation of the 30 min ocean surface elevation time series. Non-directional wave spectra were estimated from the Welch method described in "Data processing". Water velocity $U_W$ and direction $UD$ was measured with an RDI Sentinel 1200 (kHz) ADCP (0.3 m bin size) in T1 and with a Nortek Signature 1000 (kHz) ADCP in T2-3 (0.2 m bin size). Pings were ensemble averaged in 1 min intervals (The listed values are from the bin corresponding to 3 m below the surface). Wind speed $WS$ and direction $WD$ (relative to the boat) and air temperature $T_A$ was measured in 10 s intervals with a Young 81000 ultrasonic anemometer mounted 3 m above the ocean surface in T1 and T3. The listed $WD$ are absolute values obtained from the boat heading found from the GPS data. Conductivity–temperature–depth casts were performed with a SBE 19 V2 profiler to measure water density $\rho_W$, temperature $T_W$ (the listed values were measured 3 m below the surface) and salinity. The density profile presented in Fig. 4b show a strong stratification with a pycnocline around $h_0 = 10$ m below the surface. We use the going-to convention when describing direction, and the angle is defined as clockwise rotation from north. The observed oceanographic and meteorologic parameters are summarized in Table 2. It is assumed that the sea state and the wind conditions were constant over the duration of each towing experiment (~1 h, listed in Table 1), hence are the ADCP and anemometer data time averaged over this period.

Table 2. Oceanographic and meteorologic parameters.

| Exp. | $H_S$ [m] | $U_W$ [m/s] | $UD$ [°] | $WS$ [m/s] | $WD$ [°] | $T_A$ [°C] | $\rho_W$ [kg/m³] | $T_W$ [°C] |
|---|---|---|---|---|---|---|---|---|
| T1 |  | 0.12 | 168 | 4.0 | 148 | 2 | 1026 | 4.0 |
| T2 | 0.02 | 0.02 | 302 |  |  |  | 1025 | 3.5 |
| T3 | 0.12 | 0.08 | 188 | 3.8 | 68 | 0 | 1025 | 3.7 |

Wave measurements were not performed in T1, but the wave conditions were visually observed to be calm on this day, probably due to the fact that the wind was coming from south-west where the fetch was small. No considerable waves were measured in T2 and the wind conditions were visually observed to be calm on this day. Some wave activity was observed and measured in T3. The wind was coming from north-east on this day, which allowed the waves to build up in the longitudinal direction of the fjord. Deeming from the wave spectrum presented in Fig. 4a, low frequency swell coming in the larger Isfjord were dominating the wind waves.

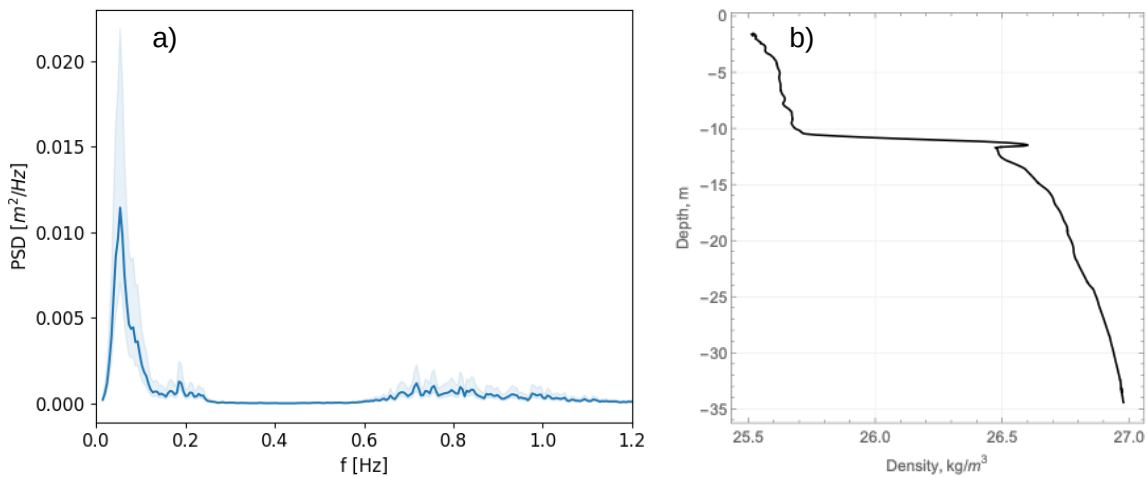

Figure 4. Oceanographic conditions. a) Wave spectrum with 95% confidence intervals during T3. Low frequency swell can be observed around 0.05 Hz and high frequency wind waves are visible around 0.8 Hz. b) Density profiles from September 23 in Tempelfjord, where the upper layer depth $h_0$ = 11 m.

**RESULTS**

Iceberg roll is defined as rotation about the IMU x-axis (randomly oriented in the horizontal plane) and yaw as rotation about the z-axis (pointing downwards). Figure 5 shows time series of iceberg roll, yaw, towing load, boat and iceberg speed and distance between boat and iceberg from T1. Three short acceleration tests were performed in the time span 14:20-14:27. A fourth acceleration test was carried out 14:30-14:34, where the towing load was increased up to 8.3 kN. The maximum towing speed was approximately 0.6 m/s. There were oscillations in the towing load with 6.1 s period, which can be seen in the inset plot of the rope tension during maximum load. The system oscillations were also present during slack conditions (when the rope was tight, but the load was small), although the amplitude was smaller. The oscillation period is further addressed in the towing load spectra in Fig. 8. Spectra from iceberg translation in the horizontal (surge/sway) and vertical (heave) direction (not presented) show oscillations with 12.8 and 7.3 s periods, respectively.

The acceleration tests shown in Fig. 5 induced an oscillating iceberg roll motion with periods of 34-51 s (from the roll spectra, not presented). Immediately after the third and fourth acceleration test, the amplitude of the roll oscillations grew quite large and was slowly damped. The iceberg yawed approximately 70° immediately after the fourth acceleration test. A slowly increasing roll motion was initiated by the third acceleration test. The roll angle increased from 3° initially to 15° at 15:20, when a sudden event, possibly a part falling off the iceberg or a shift in towline position, made the iceberg roll back to 1° with a slow damping, now with oscillation period of 26 s. A fifth acceleration test was carried out around 15:30, but the iceberg started to roll shortly afterwards, and the test had to be terminated 15:32 when the iceberg rolled around 90°. The IMU ended up floating on the surface and was barely accessible for retrieval.

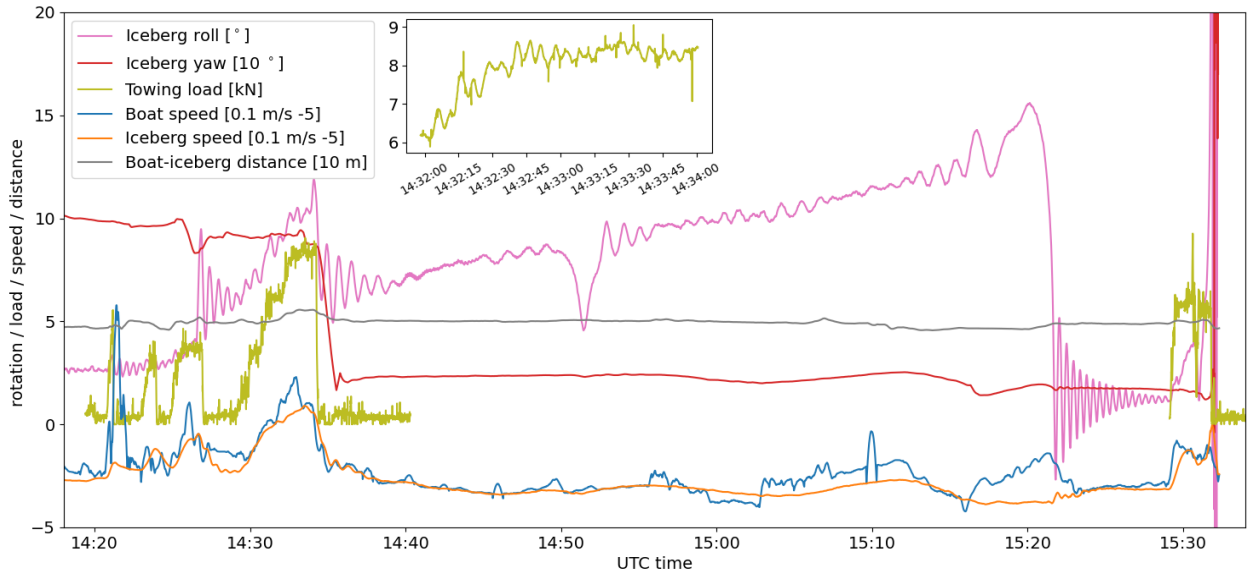

Figure 5. T1 time series of iceberg roll (pink), iceberg yaw (red), towing load (green), boat (blue) and iceberg (orange) speed and distance between boat and iceberg (gray). The speeds are multiplied by 10 and shifted 5 units down and the distance and yaw angle are divided by 10 to increase the readability. The inset plot shows a close-up of the rope tension during maximum load. There was an interruption in the experiment ~14:40-15:30 with slack conditions, during which the load cell was switched off.

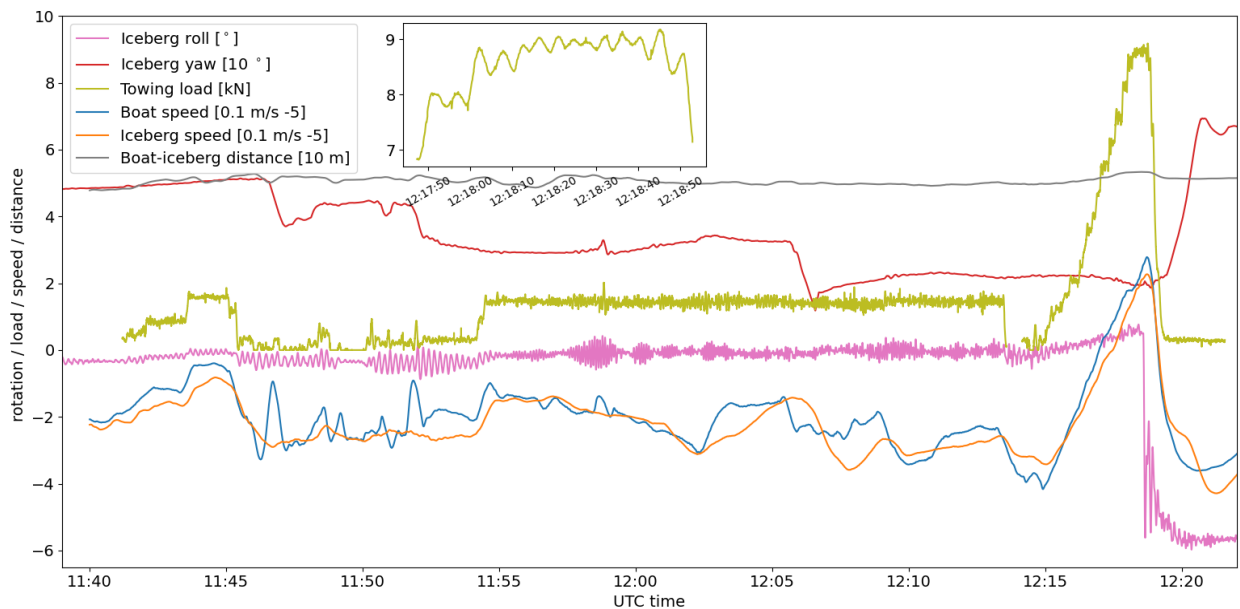

Figure 6. T2 time series. See figure text of Fig. 5 for further details.

Figure 6 shows the time series from T2. The first acceleration test was performed 11:42, during which the period of roll oscillation was 12.7 s (from the roll spectra, not presented). Thereafter, the iceberg was towed with constant motor power in a curved line in the time span 11:55-12:13, where the period of oscillation in iceberg roll decreased to 5.7 s. The maximum rate of change in iceberg course was approximately 32°/min. A second acceleration test was performed at 12:15, where the maximum towing load reached 9.0 kN, the towing load oscillation period was 6.3 s (shown in the inset plot) and the maximum boat and iceberg speed reached 0.75 m/s. The distance between the boat and the iceberg decreased from 53.3 m during maximum load to 51.3 m during slack conditions immediately after, i.e. the rope extended approximately 2 m or 4%. After the towing stopped, the iceberg rolled 6° and yawed 45° immediately after the load dropped.

Figure 7 shows the time series from T3 without the lost IMU data. Waves were present this day. The towing direction was towards the wind waves and perpendicular to the wind waves in inset plots a) and b), respectively, assuming that the wind waves were traveling in the same direction as the wind. The oscillations in towing load were more prominent in the first case, suggesting that the wave load affected the towing. Periods of oscillation were 4.5 s in both cases.

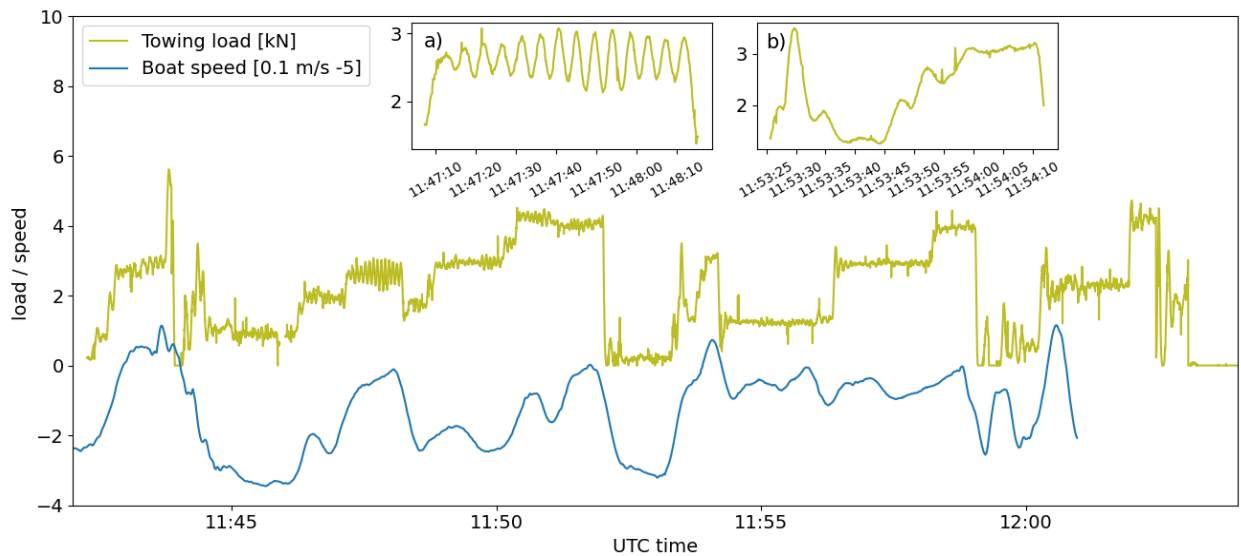

Figure 7. T3 time series. See figure text of Fig. 5 for further details. Inset plots are from instances when the boat was steaming a) against the wind waves and b) perpendicular to the wind waves.

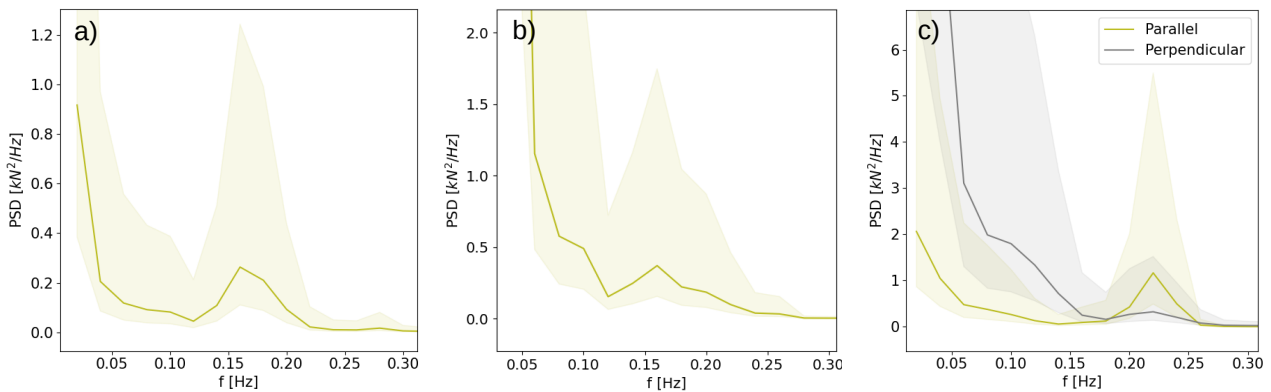

Figure 8. Towing load spectra from 2 min time series corresponding to the inset plots in Figs. 5-7 with 95% confidence intervals (~6 degrees of freedom). a) T1. b) T2. c) T3, where towing direction against (gray) and perpendicular to (green) the wind waves are shown.

Power spectral densities were calculated from 2 min time series of the towing load with the Welch method described under "Data processing", except that the segment length was 512 data points. Figure 8a-c show the towing load spectra obtained from the time series presented in the inset plots of Figs. 5-7, and the local peak frequency corresponds to periods of 6.1, 6.3 and 4.5 s, respectively. The spectra in Fig. 8c confirms that the towing load oscillations were greater when the towing direction was against (green) than when it was perpendicular to (gray) the wind waves. The increase in system oscillation amplitude may be caused by the wave drag, which influences the iceberg equation of motion in the axial direction.

The resistance of the water to the motion of the iceberg is equal to and opposite directed as the force applied by the rope during steady towing. That is when the wave and wind drag forces are neglected. The water-iceberg form drag coefficient $C_{W,I}$ can be estimated as

$$C_{W,I} = \frac{F_t}{\rho_W S_x (U_I - U_{W,x})^2}, \qquad (4)$$

where $S_x$ is the vertical cross-sectional area of the submerged part of the iceberg that is perpendicular to the axial direction, $U_I$ is the iceberg speed and $U_{W,x}$ is the water speed in the axial direction (Marchenko and Eik, 2012). Equation 4 is used to estimate $C_{W,I}$ in T1-2 when the wind and wave conditions were relatively calm, and their associated resistance forces can be neglected. $S_x$ is approximated as $h_k l$, where $l = (4S/\pi)^{0.5}$ is a representative iceberg length scale in the horizontal direction. The drag coefficient is evaluated at 14:33 in T1 and at 12:00 in T2, where the iceberg trajectory and speed and the towing force were relatively constant, and the values 0.70 and 0.54 are obtained, respectively. These values agree with Robe (1980), who reported $0.5 < C_{W,I} < 1$.

**DISCUSSION**

The observed oscillations in the towing load may be due to the towline elastic properties described in Eq. 3. The rate of change in $F_t$ with respect to $X$ can be estimated from the rope properties given by the manufacturer: 15% extension at maximum tension of 2.3 tons. Since the rope was deployed around the iceberg, the maximum tension should be doubled. With a rope length of 92/2 m, $dF_t/dX$ = 6540 N/m, which corresponds to $T_s$ = 6.5 s. Another approach is to estimate $dF_t/dX$ from the measured towing load and rope extension, i.e. 9.0 kN/2 m = 4500 N/m, which corresponds to $T_s$ = 7.8 s. At least the first estimate corresponds well with the observed oscillating period of 4.5-6.3 s. Both estimates assume a linear relation between rope tension and extension, which is a simplification of reality. It should be emphasized that the measured rope extension, i.e. the change in distance between the boat and the iceberg, was calculated from the GPS positions, which has an accuracy in the order of 1 m. This uncertainty propagates to the velocities presented in Figs. 5-7.

Towing load oscillations could also be caused by iceberg motion. The observed periods of roll oscillation were > 30 s during T1, i.e. much greater than the towing load oscillations with periods of 6.1 s. However, during T2, periods of 5.7 s were observed in roll. Natural oscillations in heave with period 7.3 and 4.7 s during T1 and T2, respectively, are also likely causes for the towing load oscillations. In the case where the towline was tightly fixed around the iceberg, small vertical movements of the iceberg around an equilibrium state may have changed the length of the rope and altered the towing load.

Large oscillations in iceberg roll were observed immediately after the towing load dropped. It is likely that the force applied by the rope tilted the iceberg during towing, and when the towing load dropped, the iceberg returned to a stable static position. Only very symmetric floating bodies have a continuum of stable positions, e.g. spheres. Normal bodies with several faces, such as icebergs, have a finite number of stable static positions. Nonlinear oscillations induced by the towing may have influenced transitions between stable positions. This could explain the sudden roll observed 15:20 in T1 and 12:18 in T2. The temperature was above the freezing point which influenced systematic disconnection of ice features from the iceberg during the towing and consequentially led to iceberg instability in the water. The non-spherical iceberg shape may prevent rotation when angular momentum is applied by the towing line, which could explain the observed oscillations in the horizontal translation. Indeed, the yaw angle was relatively constant during towing, large changes occurred immediately after the load dropped.

The stratification shown in Fig. 4b is explained by ice melting. Internal waves may be generated at the pycnocline, which could impose a large additional drag force on the iceberg. This phenomenon is known as "dead water" in ship terminology and can reduce the ship speed substantially compared to normal conditions with equal propulsion. A strong internal wave and dead water resistance force can be produced when the ratio between the ship draught and the upper layer depth $h_0$ is close to 1 (Grue, 2018). At least in T1, $h_k/h_0$ was close to unity and the dead water may have substantially increased the resistance on the iceberg.

## CONCLUSIONS

Three iceberg towing experiments were carried out on Svalbard in September 2020. The presented towing technique, which consisted of partly submerging the towing line around the iceberg, proved successful in the sense that towline slippage was avoided, and the iceberg trajectory was altered. However, the iceberg rolled approximately 90° in one situation and 180° in another. This illustrates that iceberg towing is not trivial and the importance of a well organized methodology with focus on safety. It is advantageous to have an impression of the iceberg sub-surface geometry in advance of the towing in order to be able to adjust the depth of the towline. This was obtained with an ROV in these experiments. The above-surface geometry was also documented, which enabled estimation of the iceberg mass (145-424 tons).

Boat and iceberg motion was measured with a three-axis accelerometer and gyroscope, and GPS trackers. Various met-ocean parameters were monitored, and the towline tension was measured with a load sensor. This massive instrumentation allowed for a detailed study on the dynamics of the iceberg and of the boat-rope-iceberg system. Slowly damped oscillations with period around 30 s were observed in iceberg roll, particularly right after the towing load dropped, which probably means that the iceberg returned to, or transitioned between stable static positions. Oscillations with period around 6 s were observed in the towing load. This phenomenon is attributed to either the rope elastic properties and/or the iceberg heave motion. Although the maximum significant wave height was only 12 cm, the amplitude of the towing load oscillation was significantly greater when the towing was performed against, compared with perpendicular to the wind waves, probably due to the extra wave drag on the iceberg in the axial direction. In addition to waves, upper ocean stratification could increase the iceberg drag according to the dead water phenomenon, and may be important to consider during towing operations, especially when the iceberg draft is comparable to the pycnocline depth, which was the case in one of the reported experiments.

## ACKNOWLEDGMENTS

The authors are grateful to Sebastian Sikora and Audun Tholfsen for their assistance. We also thank Nataly Marchenko for the iceberg 3D model and Malin Johansson for the satellite image. Funding for the experiment was provided by The Norwegian Society of Graduate Technical and Scientific Professionals under the Harald Boes scheme and the Research Council of Norway under the PETROMAKS2 scheme (DOFI, Grant number 28062) and the Arctic field Grant scheme (Investigation of iceberg dynamics in Isfjorden, Grant number 310691). The data are available from the corresponding author upon request.

## REFERENCES

Abramov, V.A., & Tunik, A.A., 1996. *Atlas of Arctic icebergs: the Greenland, Barents, Kara, Laptev, East-Siberian and Chukchi seas and the Arctic Basin.* Backbone Pub. Co.

Crocker, G., Wright, B., Thistle, S. & Bruneau, S., 1998. *An assessment of Current Iceberg Management Capabilities*. Contract Report for: National Research Council Canada, Prepared by C-CORE and B. Wright and Associates Ltd., C-CORE Publication 98-C26.

Earle, M.D., 1996. Nondirectional and directional wave data analysis procedures. *NDBC Tech. Doc*, 96(02).

Efimov, Y.O., Kornishin, K.A., Sochnev, O.Y., Gudoshnikov, Y.P., Nesterov, A.V., Svistunov, I.A., Maksimova, P.V., & Buzin, I.V., 2019. Iceberg Towing in Newly Formed Ice. *Proceedings of the 29$^{th}$ International Ocean and Polar Engineering Conference,* pp.911-917.

Feltham, D., 2015. Arctic sea ice reduction: the evidence, models and impacts. *Philosophical Transactions of the Royal Society of London A: Mathematical, Physical and Engineering Sciences*, 373(2045).

Grue, J., 2018. Calculating FRAM's dead water. *The Ocean in Motion*. Springer, Cham, pp.41-53.

Kornishin, K.A., Efimov, Y.O., Gudoshnikov, Y.P., Tarasov, P.A., Chernov, A.V., Svistunov, I.A., Maksimova, P.V., Buzin, I.V., & Nesterov, A.V., 2019. Iceberg Towing Experiments in the Barents and Kara seas in 2016-2017. *Proceedings of the 29$^{th}$ International Ocean and Polar Engineering Conference*, pp.905-910.

Marchenko, A., 2006. Influence of added mass effect on rotation of a drifting iceberg in non-stationary current. *Proceedings of the 14$^{th}$ International Conference on Nuclear Engineering.*

Marchenko, A., & Eik, K., 2012. Iceberg towing in open water: Mathematical modeling and analysis of model tests. *Cold Regions Science and Technology*, 73, pp.12-31.

Marchenko, A., & Gudoshnikov, Y., 2005. The influence on surcace waves on rope tension by iceberg towing. *Proceedings of the 18$^{th}$ International Conference on Port and Ocean Engineering Under Arctic Conditions,* 2, pp.543-554.

Marchenko, A., Zenkin, A., Marchenko, N., Paynter, C., Whitchelo, Y., Ellevold, T.J., & Jensen, A., 2020. Monitoring of 3D motion of drifting iceberg with an ice tracker equipped with accelerometers. *Proceedings of the 25$^{th}$ IAHR International Symposium on Ice.*

Rabault, J., Sutherland, G., Gundersen, O., & Jensen, A., 2017. Measurements of wave damping by a grease ice slick in Svalbard using off-the-shelf sensors and and open-source electronics. *Journal of Glaciology,* 63(238), pp.372-381.

Robe, R.Q., 1980. *Iceberg drift and deterioration.* In Colbeck, S. (Ed.), Dynamics of Snow and Ice Masses. Academic Press, New York, pp.211-259.

Smith, L.C., & Stephenson, S.R., 2013. New Trans-Arctic shipping routes navigable by midcentury. *Proceedings of the National Academy of Sciences*, 110(13), pp.1191-1195.

Sutherland, G., & Rabault, J., 2016. Observations of wave dispersion and attenuation in landfast ice. *J. Geophys. Res. Oceans*, 121, pp.1984-1997.

TopoSvalbard, 2021. *Topographical, digital map over Svalbard.* The Norwegian Polar Institute.